\documentclass[12pt,preprint]{aastex}
\newcommand{\bm}[1]{\mbox{\boldmath{$#1$}}}

\shorttitle{Infall of planetesimals onto growing giant planets}
\shortauthors{Shiraishi and Ida}

\begin{document}
\title{Infall of planetesimals onto growing giant planets: onset of runaway
gas accretion and metallicity of their gas envelopes}
\author{Masakazu Shiraishi and Shigeru Ida}
\affil{Department of Earth and Planetary Sciences, Tokyo Institute of Technology, 
Ookayama, Meguro-ku, Tokyo 152-8551, Japan}
\email{ida@geo.titech.ac.jp}

\begin{abstract}
We have investigated the planetesimal accretion rate 
onto giant planets that are growing through gas accretion, using 
numerical simulations and analytical arguments.
We derived the condition for gap opening 
in the planetesimal disk, which is determined by
a competition between the expansion of the planet's Hill radius 
due to the planet growth
and the damping of planetesimal eccentricity due to gas drag.
We also derived the semi-analytical 
formula for the planetesimal accretion rate
as a function of ratios of the rates of
the Hill radius expansion, the damping,
and planetesimal scattering by the planet.
The predicted low planetesimal accretion rate due to gap opening
in early gas accretion stages quantitatively shows that "phase 2," 
which is a long slow gas accretion phase before onset of 
runaway gas accretion, is not likely to occur.
In late stages, rapid Hill radius expansion 
fills the gap, resulting in significant planetesimal accretion,
which is as large as several $M_{\oplus}$ for Jupiter and Saturn.
The efficient onset of runaway gas accretion
and the late pollution may reconcile
the ubiquity of extrasolar giant planets with 
metal-rich envelopes of Jupiter and Saturn inferred from
interior structure models.
These formulae will give deep insights into formation of extrasolar gas giants 
and the diversity in metallicity of transiting gas giants.
\end{abstract}

\section{Introduction}

Models of the interior structure of Jovian planets in our solar system 
suggest that Jupiter and Saturn would contain much more amount 
of heavy elements in their envelopes than that assuming 
the solar metallicity \citep{SG04}.
This may imply that significant amount of planetesimals was accreted 
onto the planets together with gas accretion from the protoplanetary disk.
However, the orbital calculations \citep[e.g.,][]{TI97,ZL07} showed that 
the coupled effect of excitation of planetesimals' eccentricities
due to gravitational scattering by the planet
and their damping by aerodynamical and/or dynamical drag
tends to open up a gap in the planetesimal disk,
resulting in truncation of planetesimal infall 
onto the planet's gas envelope.

\citet{PO96} {\it a priori} assumed the maximally efficient 
planetesimal accretion
during gas accretion phase.  As a result of increase in the planet's mass,
the width of its feeding zone, which is proportional to 
cubic root of the mass, expands.
They assumed that planetesimals in the expanded zones are accreted
with the fastest rate for circular orbits of planetesimals.
Their assumption can be consistent with the anticipated metal-rich envelopes
of Jupiter and Saturn, but is inconsistent with the eccentricity excitation
and gap formation shown by the above orbital integrations.

Furthermore, the assumption of the maximal planetesimal accretion
results in long "phase 2" that is very inefficient gas accretion
phase before onset of runaway gas accretion.
As explained in \S 2, envelope contraction starts when core's mass
($M_{\rm c}$) becomes larger than a critical core mass ($M_{\rm c,hydro}$).
For $M_{\rm c} > M_{\rm c,hydro}$,
pressure gradient no more supports envelope
gas hydrodynamically against the increased core's gravity 
\citep{Mizuno80, IK00}.
\citet{PO96} showed that
heat generation due to the assumed planetesimal 
accretion associated with gas accretion
supports the envelope quasi-hydrodynamically (in other words,
it increases the critical core mass; eq.~[\ref{eq:crit_core_mass}]),
after the onset of envelope contraction.  
The quasi-hydrodynamical state is called ``phase 2'' 
and it may last for more than Myrs.
However, the inefficient gas accretion would be inconsistent with the
ubiquity of extrasolar giant planets \citep{IL08}.

Recently, likelihood of phase 2 is re-addressed. 
\citet{Fortier07} showed that even if the gap formation is neglected,
more realistic planetesimal accretion rate based on oligarchic growth
\citep{KI98, KI02, Thommes03} significantly suppresses 
the duration of phase 2.
\citet{ZL07} showed that planetesimals around a protoplanet
with $M_{\rm c} \sim M_{\rm c,hydro}$ 
are gravitationally shepherded and cannot be accreted.
It suggests non-existence of phase 2.
They also showed that the anticipated metal-rich envelopes
of Jupiter and Saturn is not inconsistent with it, because
such shepherding occurs only in early stages.
The planet starts accreting planetesimals 
when its mass becomes comparable to
that of gas giants
because the planetesimals trapped in mean-motion
resonances are released by resonance overlapping due to 
the planet's mass increase.
The planetesimal accretion no more halts gas accretion
onto such a massive planet.
Through numerical simulations with two different simple gas accretion 
prescriptions, they estimated that total accreted mass can be
as large as several earth masses.

The idea by \citet{ZL07} reconciled the efficient formation of
gas giants with the anticipated metal-rich envelope.
However, since they showed only numerical results with limited 
prescriptions for gas accretion, it is not clear
that the accreted planetesimal mass for more realistic gas accretion rate
is as much as that they obtained.
Furthermore, they discussed suppression of phase 2 only
qualitatively.
As shown below, the planetesimal accretion rate does depend on gas accretion
speed as well as the planet's mass.
Hence, for evaluation of amount of planetesimal infall for
more realistic gas accretion models and quantitative 
discussion on the suppression of "phase 2," general formulae
for gas accretion rate as a function of the planet's mass ($M$)
and its increase rate ($\dot{M}$) is needed.

In the present paper, through orbital integrations,
we clarify the physical mechanism to determine 
the planetesimal accretion rate and 
derive detailed semi-analytical formulae for 
the accretion rate as a function of $M$ and $\dot{M}$.
We find that the total infall mass of planetesimals can be 
as much as several earth masses even for more realistic 
gas accretion models
and quantitatively show that phase 2 is unlikely to occur. 
 
The outline of this paper is as follows.
We summarize gas accretion processes onto planets in \S 2.
The method of our calculation and initial setup is described in \S 3.
With artificial simple gas accretion models,
we clarify intrinsic physics that determines 
the planetesimals accretion rate and
derive semi-analytical formulae for the accretion rate (\S 4.1 to 4.3).
Applying the formulae to realistic gas accretion models,
we discuss the metallicity of Jupiter and Saturn 
envelope (\S 4.4).
We also discuss ``phase 2'' and find 
that phase 2 is not likely to occur (\S 4.5).
The conclusion is in \S 5.


\section{Gas Accretion Onto a Core}

As mentioned in \S 1, 
when the core mass becomes larger than a critical core mass, 
pressure gradient no more supports envelope
gas hydrodynamically against the core's gravity and
hydrostatic atmosphere does not exist.
After that, heat generation due to gas envelope
contraction itself supports the envelope against 
dynamical collapse, thus the envelope undergoes
quasi-static contraction.
The contraction allows gas inflow from the disk
into Bondi radius of the planet, so that the contraction
rate is almost equal to gas accretion rate of the planet.

Here, we briefly summarize the 
prescriptions for this process for later use.
The critical core mass depends on planetesimal
accretion rate onto the core ($\dot{M}_{\rm c}$) 
and the grain opacity ($\kappa_{\rm gr}$) associated
with the disk gas.  Based on a series of numerical models,
\citet{IK00} found that the critical core mass 
for break-down of hydrostatic atmosphere is
\begin{equation}
M_{\rm c,hydro} \simeq 
10 \left( \frac{\dot{M}_{\rm c}}{10^{-6}M_{\oplus} {\rm yr}^{-1}}\right)^{0.2-0.3}
\left( \frac{\kappa_{\rm gr}}{\kappa_{\rm gr}^{\rm P}}\right)^{0.2-0.3}M_{\oplus},
\label{eq:crit_core_mass}
\end{equation}
where $\kappa_{\rm gr}^{\rm P}$ $(\sim 1{\rm cm^2 g^{-1}})$ 
is the grain opacity given by \citet{PO85}, 
who assumed dust grains with interstellar 
abundance and size distributions.
Faster accretion and higher opacity (relatively
large $\dot{M}_{\rm c}$ and $\kappa_{\rm gr}$) 
result in a warmer planetary atmosphere and 
an enhanced pressure gradient, so $M_{\rm c,hydro}$ is larger
\citep{Stevenson82,IK00}.  

\citet{PO96} assumed the most efficient planetesimal accretion
induced by expansion of feeding zone due to increase of the
planet mass, 
the rate of which is $\sim 10^{-6}M_{\oplus} {\rm yr}^{-1}$,
for $M_{\rm c} \sim 10 M_{\oplus}$.  
When the planet with $M_{\rm c} \sim 10 M_{\oplus}$ becomes 
isolated consuming planetesimals in its feeding zone,
gas envelope starts contraction and the induced
planetesimal accretion from the expanded region of the
feeding zone increases $M_{\rm c,hydro}$ 
up to $\sim M_{\rm c}$ (eq.~[\ref{eq:crit_core_mass}])
and stall gas accretion.  This self-regulated process works 
on more than Mrys until
$M_{\rm c}$ exceeds $\sim 20 M_{\oplus}$.  
This is called ``phase 2.'' 
However, as we show in \S 4.5, 
the rate of the planetesimal accretion
induced by gas accretion is not generally large enough to 
maintain phase 2.  
Then, gas accretion dominant phase starts.

For $M_{\rm c} \sim M_{\rm c,hydro}$, 
heat generation due to planetesimal accretion
marginally equilibrates with the core's gravity.
In the quasi-static contraction stage, heat generation
due to {\it envelope contraction} 
marginally equilibrates with the gravity of the planet
with total mass $M$ (including envelope mass).
The Kelvin-Helmholtz contraction timescale 
is equivalent to planet mass increase timescale
$\tau_{\rm g,acc} = M/\dot{M}$.
Replacing $M_{\rm c}$ and $\dot{M}_{\rm c}$ by
$M$ and $M/\tau_{\rm g,acc}$ in eq.~(\ref{eq:crit_core_mass}),
$\tau_{\rm g,acc}$ is given by  
\begin{equation}
\tau_{\rm g,acc} \simeq 
10^7 \left( \frac{M}{10M_{\oplus}}\right)^{-({\rm 2.3-4})}
\left( \frac{\kappa_{\rm gr}}{\kappa_{\rm gr}^{\rm P}}\right) \; {\rm yrs}.
\label{eq:tau_g,acc}
\end{equation}
Detailed numerical simulations of quasi-static evolution of 
the gaseous envelope \citep{IK00,IG06} show consistent results
at the onset of runaway gas accretion in which the envelope and
core masses are nearly equal. 
Although \citet{PO03} suggested $\kappa_{\rm gr} 
\sim 0.01\kappa_{\rm gr}^{\rm P}$ through the numerical 
simulations of coagulation and sedimentation of dust grains in the atmosphere, 
the amount and size distribution of dust grains in the atmosphere are 
highly uncertain.
Here, we adopt the results by \citet{IG06}
with $\kappa_{\rm gr}=\kappa_{\rm gr}^{\rm P}$,
\begin{equation}
\tau_{\rm g,acc} =
10^{6.5} \left( \frac{M}{10M_{\oplus}}\right)^{-3.5}
\textrm{yrs} \, ,
\label{eq:ikoma_model}
\end{equation}
as a fiducial ``realistic'' gas accretion model.

When eq.~(\ref{eq:ikoma_model}) is extrapolated to 
large $M$ ($\ga$ 100 M$_\oplus$), it may give 
unrealistically fast supply of gas from the disk.
Hence, we limit the gas accretion rate as bellow.
\citet{TW02} showed through two-dimensional local hydrodynamic simulations, 
the mass infall to the circumplanetary subdisk 
from the protoplanetary disk is limited by
\begin{equation}
  \frac{\dot{M}}{M} \simeq 6 \times 10^{-4} f_g 
  \left( \frac{a}{5\textrm{AU}} \right)^{-1.5} 
  \left( \frac{M}{M_{\oplus}} \right)^{0.3} \textrm{yr}^{-1} \, ,
  \label{eq:tanigawa_model}
\end{equation}
where $f_g$ is a scaling factor for disk gas surface density
defined by eq.~(\ref{eq:surface_gas_density}).
We use this limit with $f_g = 0.7$.

\begin{figure}[btp]
  \epsscale{0.6}	%
  \plotone{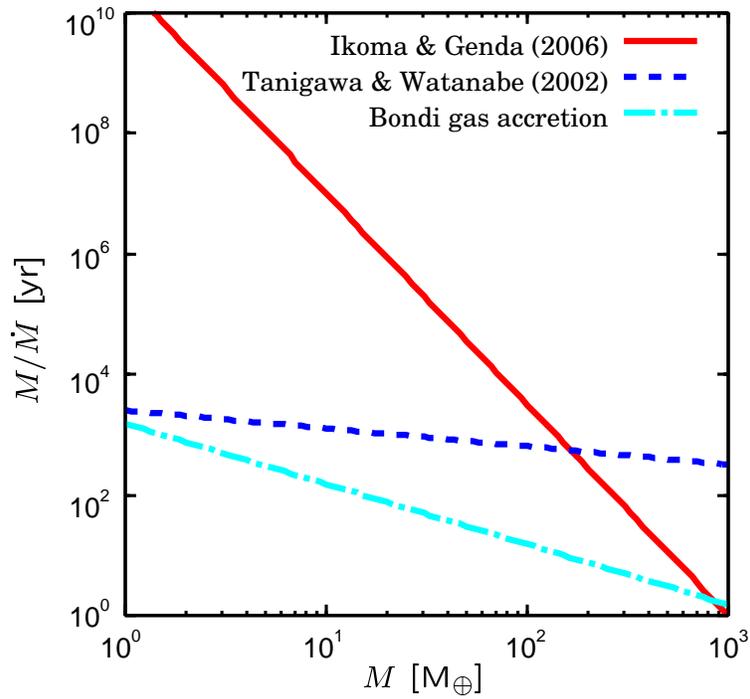}	%
  \caption{Gas accretion timescales for a planet with mass $M$
at $a = 5 \textrm{AU}$.
Solid line is an extrapolation of the model by \citet{IG06}.
Dashed and dotted-dashed lines represent limits by \citet{TW02} and
Bondi accretion.
}
  \label{fig:Increase_timescale}
\end{figure}

Another limit is Bondi gas accretion, the rate of which
is given by $\dot{M} = \pi r^2_B \rho_{gas} c_s$,
where $\rho_{gas} = \Sigma_{g}/(2H)$ is the spatial 
density of gas disk and $H$ is the disk scale hight, 
$r_B = 2GM/c_s^2$ is the Bondi radius, 
$c_s = H\Omega$ is the sound speed and 
$\Omega = \sqrt{GM_{\odot}/a^3}$ 
is the Keplerian angular velocity.
Adopting the temperature distribution in the limit of 
an optically thin disk (Hayashi 1981), 
$T = 2.8\times 10^2 (r/\textrm{1AU})^{-1/2} \, \textrm{K}$, 
the Bondi gas accretion limit is
\begin{equation}
 \frac{\dot{M}}{M} = 0.7 \times 10^{-3} 
 \left( \frac{a}{\textrm{5AU}} \right)^{-2} 
 \left( \frac{M}{M_{\oplus}} \right) \, \textrm{yr}^{-1} \, .
 \label{eq:bondi_model3}
\end{equation}
As figure \ref{fig:Increase_timescale} shows,
the timescale ($=M/\dot{M}$) in eq.~(\ref{eq:tanigawa_model}) 
is generally longer than the Bondi accretion timescale, 
so an actual lower limit for gas accretion timescale 
is given by eq.~(\ref{eq:tanigawa_model}).

\section{Calculation Setup}
\subsection{Orbital Integration}
We numerically calculate the orbital evolution of 
a swarm of planetesimals in the vicinity of a protoplanet's orbit 
embedded in a gaseous disk.
The protoplanet grows accreting gas with a given rate.
The planetesimals are treated as massless test particles
and neglect their interactions. 
We assume that the protoplanet has a fixed circular orbit.

The planetesimals' orbits are affected by the gravitational force
of the growing  protoplanet and drag force from disk gas, 
\begin{equation}
  \bm{f}_{\rm gas} = - \frac{\bm{v} - \bm{v}_{\rm gas}}{\tau_{\rm damp}}\, ,
  \label{eq:gas_drag}
\end{equation}
where $\bm{v}$ and $\bm{v}_{\rm gas}$ are the velocity of a planetesimal 
and disk gas.  
The gas motion is a circular Keplerian motion.
In some runs, we adopted slightly slower rotation speed of
the disk gas due to radial pressure gradient in disk gas
\citep[e.g.,][]{AD76}, which induces inward migration of
planetesimal orbits.
However, we found that the inward migration hardly changed 
the results of planetesimal accretion rate onto the protoplanet.
We here show the results without the inward migration. 
We set a damping timescale of the gas drag, 
$\tau_{\rm damp}$ $(= e/\dot{e})$, 
as a constant parameter for all the planetesimals throughout a run 
in order to make clear the effect of the damping force.

We follow the prescription of gas surface density distribution
by \citet{IL04}, 
\begin{equation}
  \Sigma_g = 210 f_g \left( \frac{a}{\textrm{5AU}} \right)^{-3/2} \ \textrm{g$\,$cm$^{-2}$} \, ,
  \label{eq:surface_gas_density}
\end{equation}
where $f_g$ is the scaling parameter and 
$f_g = 0.7$ corresponds to the gas surface density 
of the minimum mass solar nebular model, $\Sigma_{g,{\rm MMSN}}$ \citep{HA81}.
For simplicity, we neglect a gap in the gas disk, which may
be opened up by the perturbations from a massive 
protoplanet \citep[e.g.,][]{LP93} 
and assume the above unperturbed $\Sigma_g$ everywhere.     
With this $\Sigma_g$, a given value of $\tau_{\rm damp}$ corresponds to 
individual planetesimal mass \citep{AD76,TI99},
\begin{equation}
  m = 3 \times 10^{17} f_g^3 \left( \frac{e}{0.1} \right)^3 
  \left( \frac{\tau_{\rm damp}}{10^5\textrm{yr}} \right)^3
  \left( \frac{\rho_{\rm pl}}{1\textrm{g$\,$cm$^{-3}$}} \right)^{-2} 
  \left( \frac{a}{\textrm{5AU}} \right)^{-39/4} \ \textrm{g} \, ,
\label{eq:m_pl_gas}
\end{equation}
where $a$, $e$, and $\rho_{\rm pl}$ are semi-major axis, 
eccentricity, and material density of planetesimals, respectively.
If gravitational drag \citep[e.g.,][]{TW04} is considered
in stead of aerodynamical gas drag,
\begin{equation}
  m = 4.5 \times 10^{26} f_g^{-1} 
  \left( \frac{\tau_{\rm damp}}{10^5\textrm{yr}} \right)^{-1}
  \left( \frac{a}{\textrm{5AU}} \right)^{2} \ \textrm{g} \, ,
\label{eq:m_pl_gas2}
\end{equation}
although in this case, interactions among the planetesimals
could be important.

The orbits of planetesimals are numerically integrated by using the fourth-order Hermite scheme \citep{MA92} with the hierarchical timestep \citep{MA91}.
The equation of motion of particle $k$ is given by
\begin{equation}
  \frac{d^2\bm{r}_k}{dt^2} = -GM_\odot \frac{\bm{r}_k}{|\bm{r}_k|^3} 
- \frac{GM (\bm{r}_k - \bm{r})}{|\bm{r}_k - \bm{r}|^3} 
- \frac{GM \bm{r}}{|\bm{r}|^3} + \bm{f}_{\rm gas}\, ,
  \label{eq:planetesimals}
\end{equation}
where $M$ and $\bm{r}$ is the mass and position of the protoplanet.
The first term to the last one represent the gravity from the central star, 
the gravitational perturbation from the protoplanet, 
the indirect term and the gas drag force, respectively.
We set $M_* = M_\odot$.

When a planetesimal contacts the surface of the protoplanet, 
the planetesimal is removed after recording the collision.
The planet mass is unchanged.
The physical radius of a protoplanet is 
determined by its mass and internal density $\rho$ as
\begin{equation}
  R = \left( \frac{3M}{4\pi \rho} \right)^{1/3},
  \label{eq:physical_radius}
\end{equation}
We set $\rho = 1$gcm$^{-3}$ in all simulations.
The dependence of the planetesimal accretion rates 
on $\rho$ will be discussed in \S 4.4.

Although we neglect gravitational forces of planetesimals,
mass of planetesimals is specified in order to calculate
the amount of mass accretion onto the protoplanet
(regarding ``effective'' mass for gas drag force, see below).
Assuming equal-mass planetesimals, 
they are initially distributed in the range 
$a_{\rm in} < a < a_{\rm out}$ to satisfy the surface mass density
\begin{equation}
  \Sigma_d = 3.8 f_d \left( \frac{a}{\textrm{5AU}} \right)^{-3/2} \textrm{g}\,\textrm{cm}^{-2}\, ,
  \label{eq:surface_mass_density}
\end{equation}
where $f_d$ is a scaling factor. As is the case for $f_g$,
$f_d = 0.7$ corresponds to MMSN.
The inner and outer boundaries are 
$a_{\rm in} = a_p(1 - 5 h_{\rm f})$ and 
$a_{\rm out} = a_p(1 + 10 h_{\rm f})$, 
where $a_p$ is the semi-major axis of the protoplanet and $h_{\rm f}$ 
is the reduced Hill radius for final mass of the planet ($M_{\rm f}$).
The reduced Hill radius of a protoplanet $h$ is
Hill radius $r_{\rm H}$ scaled by $a_p$, 
\begin{equation}
  h = r_{\rm H}/a_p = \left( \frac{M}{3M_*} \right)^{1/3} \, .
  \label{eq:reduced_Hill_radius}
\end{equation}

In all numerical simulations, we adopt
$a$ = 5AU, $M_{\rm f} = M_J$ (Jupiter mass), and
$f_d = f_g$.
Accordingly, $h_{\rm f} = 6.8 \times 10^{-2}$,
$a_{\rm in} = 3.3$ AU, and $a_{\rm out} = 8.4$AU.
We will derive the dependences on $a$, $M_{\rm f}$, and $f_d (= f_g)$
by analytical arguments and discuss the results with
different parameter values.
Total mass of planetesimals within the region 
$a_{\rm in} < a < a_{\rm out}$ is 
$\sim 20 f_d M_\oplus$.
The number of planetesimals in most runs is $N = 20000$. 
With an assumption that planetesimals have an equal mass, their
individual mass is 
$m_{pl} \simeq 1.1 \times 10^{-3} f_d M_\oplus 
= 6.6 \times 10^{24} f_d$ g.
In our simulations, we specify
$\tau_{\rm damp} = 10^6, 10^5, 10^4$ yrs and $\infty$ (gas-free case), 
independent of the values of $m_{pl}$. 
Except for the gas-free case, 
the above values of $m_{pl}$ is much larger than the 
values in eq.~(\ref{eq:m_pl_gas}) for the given $\tau_{\rm damp}$,
so planetesimals that we use correspond to ``super particles''
representing many smaller planetesimals.
Since we neglect interactions among planetesimals,
such ``super particles'' treatment is not inconsistent.
Initial eccentricity and inclination of planetesimals are taken as $e_0$ = $i_0$ = 0.001 for all simulations.
Since $e$ and $i$ are quickly pumped up by perturbations from the protoplanet, the choice of $e_0$ and $i_0$ does not affect results.

\subsection{Growth of a Protoplanet}

Since we consider the phase after isolation of protoplanets,
we assume that the growth of the protoplanets 
is dominated by accretion of surrounding disk gas
but not by planetesimals.
As we will show later,
this assumption is valid, because amount of
accreted gas is much larger than the anticipated 
amount of accreted planetesimals.

In \S 2, we described the prescription for gas accretion.
In the numerical simulations,
we use simple artificial gas accretion models
in order to make clear what conditions regulate the planetesimal
accretion rate.  From the results with the artificial models,
we derive semi-analytical formulae for the planetesimal
accretion in general forms (\S 4.3).
Applying the formulae to the more realistic gas accretion
rate in \S 2, we calculate the total mass of planetesimal infall 
into the envelope of Jupiter and Saturn in \S 4.4. 

\begin{figure}[tbp]
  \epsscale{0.6}	
  \plotone{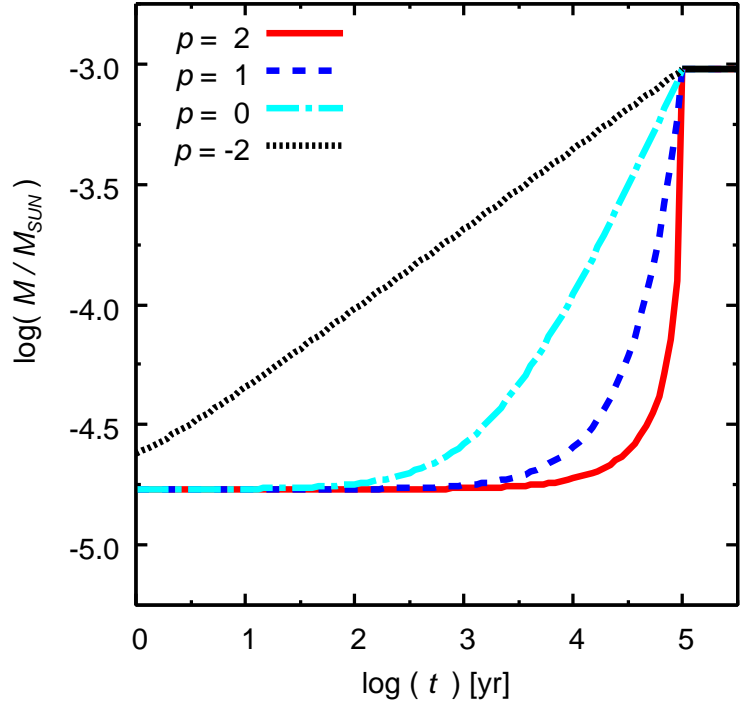}	
  \caption{Evolution of the protoplanet mass 
  according to the simple power-law gas accretion models 
  ($\dot{M} \propto M^{p}$).
  Initial and final masses are $M_0 = 5.67 M_{\oplus}$ and
  $M_{\rm f} = M_J$, where $M_J$ ($= 10^{-3}M_{\odot}$) is a Jupiter mass.
  Growth timescale $t_{\rm f} = 10^5$ yr.
  After $t > t_{\rm f}$, we set $M = M_{\rm f} = const$.}
  \label{fig:Artificial_massup}
\end{figure}

The simple artificial gas accretion models are expressed by
\begin{equation}
  \frac{dM}{dt} \equiv \alpha M^p \, ,
  \label{eq:arti_model}
\end{equation}
where $\alpha$ is the integration constant determined by 
the boundary condition. 
We set the condition as $M = M_0$ for $t = 0$ and 
$M = M_{\rm f}$ for $t = t_{\rm f}$.
Following \citet{ZL07}, we set the protoplanet at $5$AU 
with its initial mass $M_0 = 5.67 M_\oplus$ and 
final mass $M_{\rm f} = M_J$ (Jupiter mass).
We adopt $t_{\rm f} = 10^5 $yr for numerical simulation,
following their nominal cases.
The growth with $p = 2$ and 0 correspond to the Bondi and 
linear models
in \citet{ZL07}.
Figure \ref{fig:Artificial_massup} shows the evolution of 
the mass of a protoplanet by accreting gas 
for $p = 2, 1, 0, -2$.
Here we assume that $M_0 > M_{\rm c,hydro}$.
The consistency of the assumption is checked in \S 4.5.  

\section{Results of Orbital Calculation}
\label{sec:result}

\subsection{Overall Evolution}

\begin{figure}[htbp]
 \epsscale{.60}   %
  \plotone{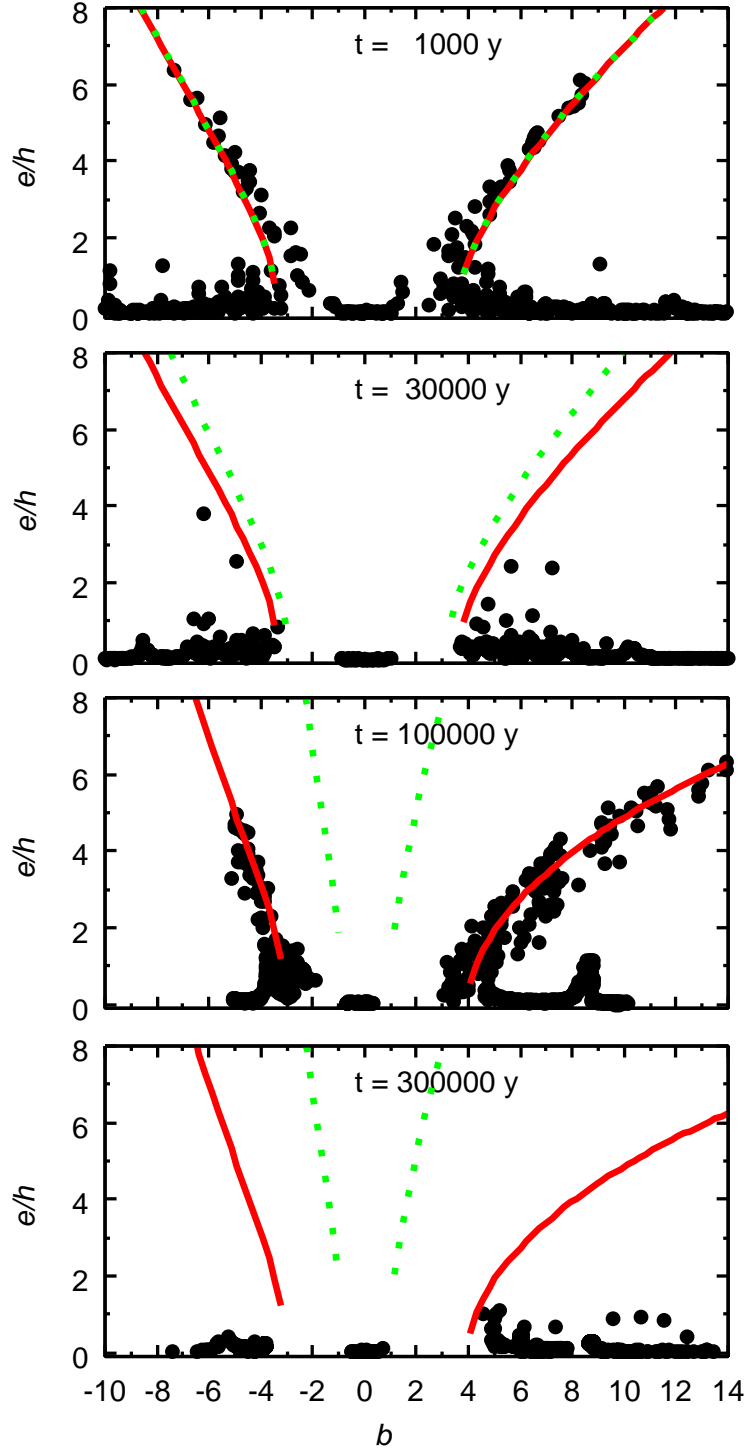}
  \caption{Orbital evolution of
a swarm of a planetesimals on the $b$-$(e/h)$ plane.
We adopt $p = 2$, $\tau_{\rm damp} = 10^4$yrs 
and $t_{\rm f} = 10^5$ yrs.
The planet is fixed at $e/h = b = 0$ ($a_p = $5 AU).
The horizontal axis $b$ expresses 
$(a-a_p)/h$ where $a$ is the semimajor axis of planetesimals.
Solid and dotted lines represent 
the boundaries of the feeding zone (i.e. Jacobi energy $E_J = 0$) 
and those at $t = 0$, respectively.
The time evolution of the latter is caused by
increase in $h$.
The selected number of planetesimals is 1000 in 3.3 AU $\leq a \leq$ 8.4 AU
at $t=0$.
The numbers of planetesimals are  998($10^3$yr), 992($3\times 10^4$yr), 904($10^5$yr), and 878($3\times 10^5$yr).}
  \label{fig:Orbit_snapshot}
\end{figure}

Figure \ref{fig:Orbit_snapshot} shows the snapshots of 
the distributions of planetesimals on the $b$-$e/h$ plane,
where $h$ is defined by eq.~(\ref{eq:reduced_Hill_radius}).
The scaled orbital separation $b$ is defined by
\begin{equation}
  b = \frac{a - a_p}{ha_p} \, ,
  \label{eq:barameter}
\end{equation}
where $a_p$ is semimajor axis of the protoplanet.
The protoplanet is fixed at the origin (i.e., $b = 0$ and $e = 0$).
The growth rate of the protoplanet $\dot{M}$ is $\propto M^2$.
To avoid busy plots, we show only 1000 planetesimals
in this figure.
We integrate the evolution of planetesimals 
for $3 \times 10^5$yrs.
Since $t_{\rm f} = 10^5$yrs,
we set $\dot{M}=0$ for $t=1$--$3 \times 10^5$yrs, which
corresponds to termination of gas accretion due to gap formation
in the gas disk, although we neglect the effect of
gas density depletion on drag force.
The damping timescale is $\tau_{\rm damp} = 10^4$yr. 

In figure \ref{fig:Orbit_snapshot}, 
we also drew the Jacobi energy $E_J$,
\begin{equation}
  E_J = \frac{1}{2}((e/h)^2+(i/h)^2) - \frac{3}{8}b^2 + \frac{9}{2} + O(h).
  \label{eq:jacobi_energy}
\end{equation}
In the figure, we also include higher order terms of $h$ in $E_J$. 
In the circular restricted three-body problem, $E_J$ is conserved 
between before and after scattering by the protoplanet (on average,
both ($e^2+i^2$) and $b^2$ increase).
Since only planetesimals with $E_J \geq 0 $ 
can enter the Hill sphere of the protoplanet \citep[e.g.,][]{HA77},
we regard the region $E_J > 0$ as the feeding zone of the protoplanet.
When $e/h, i/h \la 1$, the width of feeding zone is
$b \simeq 2\sqrt{3}$.

In the top panel ($t=1000$yrs), the planetesimals in the vicinity of 
the protoplanet are scattered and their $e$ and $b$ increase 
along a constant $E_J$ curve.
The planetesimal eccentricities $e$ 
are damped in the panel of $t = 3\times 10^4$yrs
because of $\tau_{\rm damp} = 10^4$yrs.
Since the gas drag damps $e$ keeping $b$ almost constant, 
all the planetesimals except for those trapped in horseshoe orbits
go out of the feeding zone. 
As the protoplanet grows up, its feeding zone expands.
Since $b \propto M^{-1/3}$,
$b$ of planetesimals decreases with time, but
scattering opposes it. 
Since $\dot{M} \propto M^2$, the expansion accelerates with time.
Eventually, the expansion overwhelms the gap opening due to a coupling
effect of scattering and gas drag damping, so that
planetesimals go into the feeding zone in the
panel of $t = 1\times 10^5$yrs.
After $t = 1\times 10^5$yrs, the increase of $M$ stopped, 
so a gap in the planetesimal disk is again produced (the bottom panel).
Planetesimals that have sufficiently large $b$ are
captured in proper mean motion resonances.
Although we neglect inward migration due to slightly
slower rotation of gas than Keplerian rotation due to pressure
gradient, damping of $e$ results in small decrease in $b$
due to angular momentum conservation.
Such inward migration causes resonance trapping. 
The evolution of the gap width is consistent with the
result by \citet{ZL07}. 

Thus, after initial relaxation, planetesimals are shepherded
and planetesimal accretion rate is very low, until
efficient planetesimal accretion re-starts in late stage.
\citet{ZL07} showed through orbital simulation of planetesimals
that planetesimal accretion occurs only in late stage of gas accretion.
They suggested that 
in late stage, planet's mass becomes large and 
the mean motion resonances overlap to release planetesimals
captured in the resonances.
So, they concluded that mass of the protoplanet controls
the accretion rate of planetesimals onto the protoplanet.
This effect indeed determines supply of planetesimals to
the regions near the feeding zone.
However, whether the planetesimals near the feeding zone
are shepherded or not is determined by gap opening
that is a result of competing processes of the feeding zone
expansion and scattering/damping, so 
the values of $\dot{M}$ play an important role as well as $M$.

\subsection{Dependence of Planetesimal Accretion Rate on Planet's Mass}
\label{subsec:evolution_of_planetesimals_accretion_rate}

\begin{figure}[tbp]
  \epsscale{0.8}	%
  \plotone{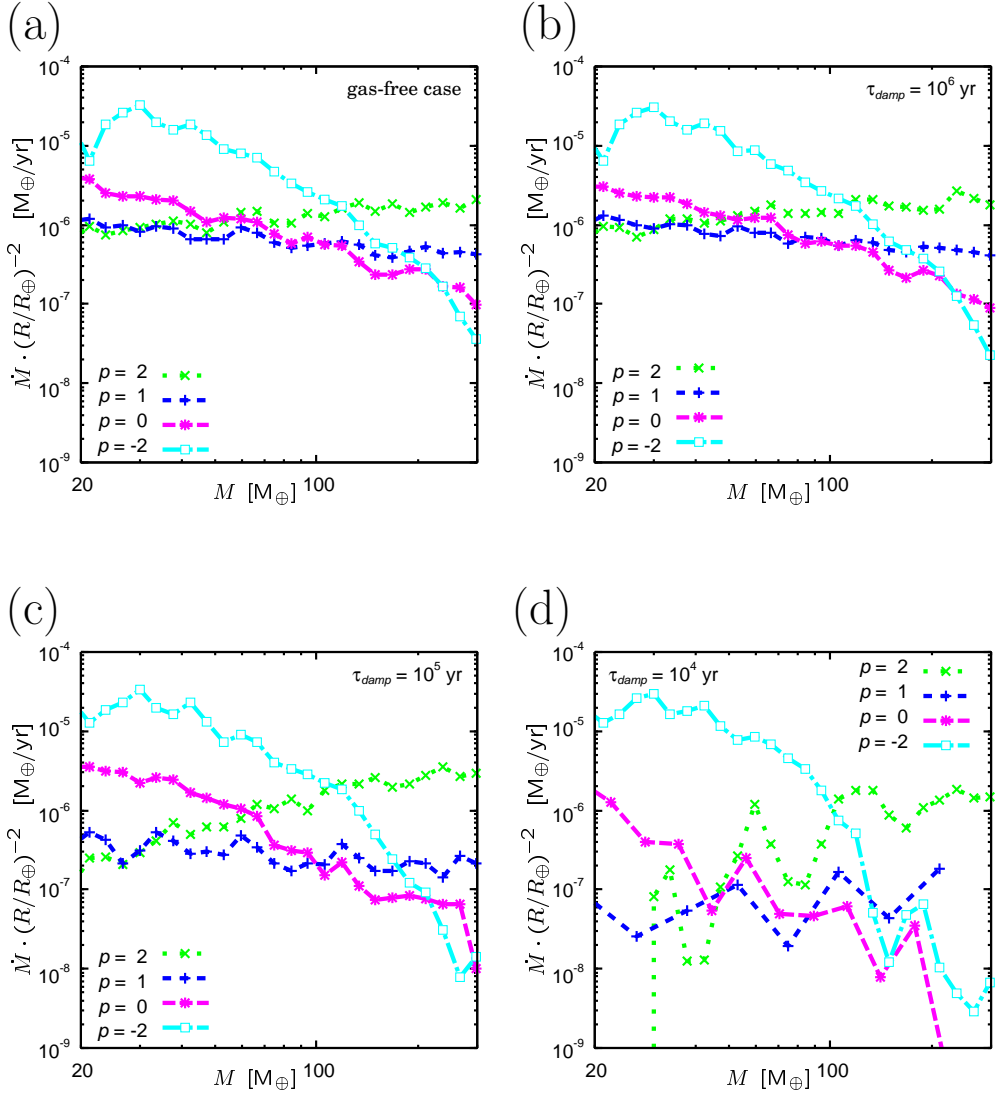}
  \caption{Evolution of the planetesimal accretion rate onto the 
growing planet as a function of the protoplanet mass ($M$).
(a) The results in the gas-free case, 
the cases of (b) $\tau_{\rm damp} = 10^6$ yrs, (c) $10^5$ yrs 
and (d) $10^4$ yrs. 
The four lines in the each panel represent the results
with various gas accretion models ($p = 2, 1, 0, -2$).
Initial mass of the protoplanet $M_0$ is set as 5.67$M_\oplus$.
The systems initially consist of 20,000 planetesimals, 
so the individual planetesimal masses 
correspond to $\simeq 7.7 \times 10^{-4}M_{\oplus}$.}
  \label{fig:mlmp7_mass}
\end{figure}

The evolution of the planetesimal accretion rate 
for $\tau_{\rm damp} = 10^6, 10^5, 10^4$ yrs and gas-free case 
is plotted as a function of $M$ in Figure \ref{fig:mlmp7_mass}.
The four lines in each panel represent various 
gas accretion models ($p = 2, 1, 0, -2$).
Initial mass of the protoplanet $M_0$ is set as 5.67 M$_\oplus$.
Starting with 20,000 planetesimals 
(i.e., planetesimal mass $\sim 1.1 \times 10^{-3} f_dM_{\oplus}$), 
we calculated for $10^5$yrs 
(the growth timescale $t_{\rm f} = 10^5$yrs).

To see the dependence on $M$ more clearly, 
we plot the scaled planetesimal accretion rate $\dot{M}/R^2$,
where $R$ is the physical radius of the protoplanet.
Through the numerical simulations, we found that planetesimals are 
likely to experience 2-D accretion rather than 3-D.
The 2-dimensional accretion rate is
\begin{equation}
  \frac{dM}{dt} \sim 2R\Sigma_d \left( \frac{v_{\rm esc}}{v_{\rm rel}} \right) v_{\rm rel} = \sqrt{\frac{32\pi \textrm{G}\rho}{3}}\Sigma_d R^2, 
  \label{eq:size-independent_accretion_rate}
\end{equation}
where $\Sigma_d$, $v_{\rm esc} = \sqrt{2\textrm{G}M/R}$ and $v_{\rm rel}$ are 
the surface density of the planetesimals, escape velocity from the 
protoplanet's surface and relative velocity between the protoplanet 
and planetesimals, respectively. 
The scaled accretion rate $(\dot{M}/R^2)$ is 
determined by effective $\Sigma_d$ in the feeding zone
for fixed internal density of the protoplanet ($\rho$).
In our simulations, the total mass of the planetesimals is not
significantly decreased, so the effective $\Sigma_d$ is
determined by scattering by the planet, gas drag, and
Hill radius expansion due to the planet growth.

Figure \ref{fig:mlmp7_mass} shows that the 
planetesimal accretion rates for $f_d = 0.7$.
The planetesimal accretion rates as a function of $M$ 
depends on the parameter $p$.
For $p = -2$ and 0, the scaled planetesimal accretion rate 
decreases with $M$, which suggests that a gap in the
planetesimal disk is formed when $M$ becomes large. 
On the other hand,
for $p=2$, the protoplanet may grow so fast in the late stage 
that the feeding zone expansion overwhelms the gap formation,
as shown in figure \ref{fig:Orbit_snapshot}.
The dependence on $p$ implies that
the accretion rate is not a function solely of $M$,
but depends on $\dot{M}$ as well as $M$,
because $p$ determines the $\dot{M}$--$M$ relation.

\subsection{Dependence of Planetesimal Accretion Rate on 
Gap Formation Parameters}

Here we show that competition among
the feeding zone expansion, scattering, and eccentricity
damping regulates flux of planetesimals across the boundaries of
the feeding zone ($E_J = 0$), that is, the planetesimal accretion rate.
We consider change rates of $b^2$ and $(e/h)^2$ of planetesimals
(we neglect the contribution from $i$ because $i$ is usually correlated 
to $e$ and $i < e$).

Evolution of planetesimals on the $b^2$--$(e/h)^2$ space
due to gravitational scattering by the protoplanet, damping of 
eccentricity by gas drag, and expansion of Hill radius by
mass increase of the protoplanet is expressed by
the change rates, $v_{\rm scat}$, $v_{\rm damp}$, 
and $v_{\rm H}$, on the plane.
Since $b \propto h^{-1} \propto M^{-1/3}$,
\begin{equation}
v_{\rm H} \equiv \frac{d b^2}{dt}{\rm (growth)} 
          = (bh)^2 \frac{d h^{-2}}{dt} 
          = -\frac{2}{3} b^2 \frac{\dot{M}}{M}
          \simeq - \frac{8}{\tau_{\rm g,acc}},
  \label{eq:hill_velocity}
\end{equation}
where $\tau_{\rm g,acc} = M/\dot{M}$ is the timescale
of planet mass increase. 
In the last equation, we used $b \simeq 2\sqrt{3}$, which is
the location of the feeding zone for $e/h \la 1$, for simplicity.
With $\tau_{\rm damp} = e/\dot{e}$,
\begin{equation}
v_{\rm damp} \equiv \frac{1}{2} \frac{d (e/h)^2}{dt}{\rm (damping)}  
             = - \frac{(e/h)^2}{\tau_{\rm damp}}. 
\end{equation}
The factor $(1/2)$ in the definition is added for 
the more simple form of the final expression and
better fit with numerical results.
Evolution due to the scattering is increase of $b^2$ and
$(e/h)^2$ on average, along a constant $E_J$ curve ($E_J \sim 0$).
The corresponding change rate is
\begin{equation}
v_{\rm scat} \equiv \frac{d b^2}{dt}({\rm scattering})
             =\frac{4}{3}\frac{d (e/h)^2}{dt}({\rm scattering}).
\end{equation}
Assuming long-range gravitational interaction with $(e/h) \la 1$, 
linear calculation \citep{GT82,HN90} showed that 
$b$ of a planetesimal is increases by 
$\delta b \simeq 30 b^{-5}$ during each encounter.
Numerical calculation showed that for $b \sim 3$--4, 
$\delta b$ is overestimated by a factor $\sim 10$ \citep{ID90}.
Since the scattering occurs at every synodic time 
$T_{\rm syn} \simeq 2\pi a_p/(\frac{3}{2} b r_{\rm H} \Omega_K)$, 
\begin{equation}
v_{\rm scat} = 2b\frac{d b}{dt}({\rm scattering})
             \simeq 2 b \frac{0.1\delta b}{T_{\rm syn}} 
             \simeq \frac{6}{b^4} \frac{(3/2)bh}{T_{\rm K}} 
             \simeq 0.22 \frac{h}{T_{\rm K}}, 
  \label{eq:scattering_velocity}
\end{equation}
where $T_{\rm K}=2\pi/\Omega_{\rm K}$ is Keplerian period and
$b \simeq 2\sqrt{3}$ is again used.

Since the feeding zone is determined by the values of
$E_J$ and the scattering does not change the values,
the condition of gap opening would be $v_{\rm damp} \ga v_{\rm H}$.
If inward migrations of planetesimals due to gas drag or 
type I migration of the protoplanet is considered
but growth of a protoplanet is neglected, 
the gap formation condition is similarly derived, 
replacing $v_{\rm H}$ by $db^2/dt$ due to
gas drag \citep{TI97} or type I migration \citep{TI99}.
These effects can suppress growth of the protoplanets
before they attain their isolation masses \citep{TI97,TI99}.

When $v_{\rm damp} \la v_{\rm H}$, the gap is not created
and planetesimals are engulfed by the expanding feeding zone.
The engulfment rate would be determined by
$v_{\rm H}/v_{\rm scat}$, because $v_{\rm H}$ and 
$/v_{\rm scat}$ have opposite
directions to each other in the $b^2$ components.
Thus, it is expected that for $v_{\rm damp} < v_{\rm H}$, 
the accretion rate would be regulated by 
\begin{equation}
  \xi \equiv | \frac{v_{\rm H}}{v_{\rm scat}} | 
       \simeq 37 h^{-1} \frac{T_{\rm K}}{\tau_{\rm g,acc}}
       \simeq 4.1 
             \left(\frac{a_p}{5\textrm{AU}} \right)^{3/2} 
             \left( \frac{M}{M_{\oplus}} \right)^{-1/3} 
             \left( \frac{\tau_{\rm g,acc}}{10^{4} \textrm{yrs}} \right)^{-1} \, ,
  \label{eq:xi}
\end{equation}
while for $v_{\rm damp} > v_{\rm H}$, 
the accretion rate would be regulated by 
\begin{equation}
  \eta \equiv | \frac{v_{\rm H}}{v_{\rm damp}} | 
       \simeq \frac{8}{(e/h)^2} \frac{\tau_{\rm damp}}{\tau_{\rm g,acc}} 
       \simeq 0.8 
       \left( \frac{\tau_{\rm damp}}{10^{4} \textrm{yrs}} \right)^{1/2}
       \left( \frac{\tau_{\rm g,acc}}{10^{4} \textrm{yrs}} \right)^{-1} 
       \left( \frac{M}{M_\oplus}\right)^{-1/6} 
       \left(\frac{a_p}{5\textrm{AU}} \right)^{3/4}, 
  \label{eq:eta}
\end{equation}
where we used eq.~(\ref{eq:equili_e}) in Appendix for $(e/h)^2$.
Since $\tau_{\rm g,acc}=M/\dot{M}$ and 
$\tau_{\rm damp}$ and $h$ are functions of $M$, 
the planetesimal accretion rate would depend on
$\dot{M}$ as well as $M$. 
We show that the numerical results agree with the above
argument and derive formulae for 
the planetesimal accretion rate as a function of $\xi$ and $\eta$.

\begin{figure}[tbp]
  \epsscale{0.8}	%
  \plotone{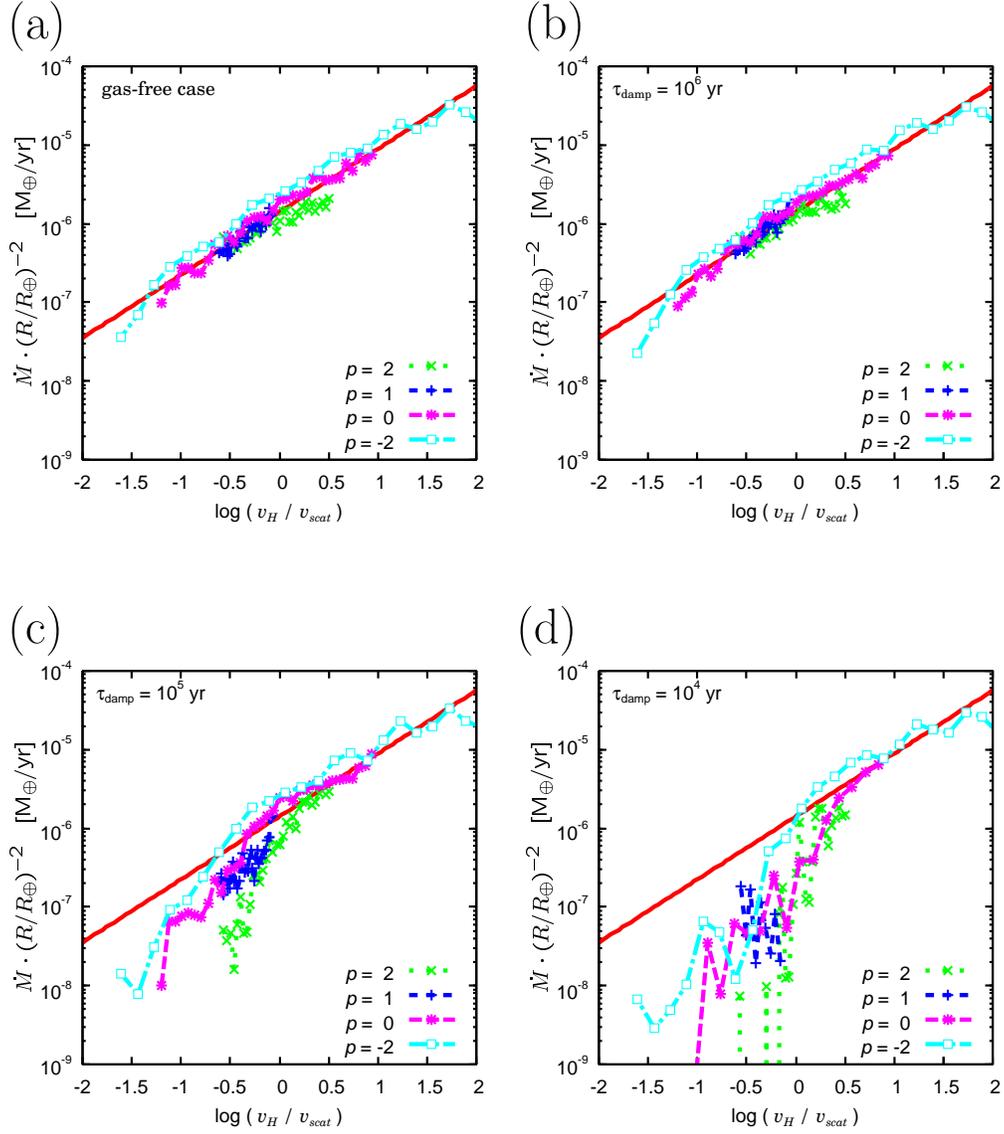}
  \caption{Evolution of the planetesimal accretion rate 
as a function of $\xi = v_H / v_{\rm scat}$ for $p = 2,1,0$ and $-2$.
(a) The results in the gas-free case, 
the cases of (b) $\tau_{\rm damp} = 10^6$ yrs, (c) $10^5$ yrs 
and (d) $10^4$ yrs. 
The fitting formula, eq.~(\ref{eq:fitting_line1}),
is expressed by thick solid lines in the plots.}
  \label{fig:mlmp7_xsi}
\end{figure}

Figure \ref{fig:mlmp7_xsi} shows the evolution of 
the scaled planetesimal accretion rate as 
a function of $\xi = v_{\rm H}/v_{\rm scat}$ 
for $p=2,1,0$ and $-2$ in the cases of
$\tau_{\rm damp} = 10^6, 10^5, 10^4$ yrs and gas-free case.
As suggested in the above discussion,
figure \ref{fig:mlmp7_xsi} shows that 
in the ranges of $\eta > 1$ 
[equivalently, $\xi > 5(\tau_{\rm damp}/10^5\textrm{yr})^{-1/2}
(M/M_{\oplus})^{-1/6}(a_p/5\textrm{AU})^{-3/4}$], 
the scaled planetesimal accretion rate is independent 
of planet gas accretion 
models with different $p$ (different $\dot{M}$--$M$ relations) and
different $\tau_{\rm damp}$.
This confirms that the accretion rate is
determined by $\xi$ for $\eta>1$ (non-gap cases).
The planetesimal accretion rate in this case is 
given by
\begin{equation}
  \frac{dM_{\rm solid}}{dt} =
  10^{\beta} 
  \left( \frac{R}{R_\oplus}\right)^{2} 
  f_d \left( \frac{v_{\rm H}}{v_{\rm scat}} \right)^{\alpha} 
  \, \textrm{M}_\oplus \, \textrm{yr}^{-1} \, ,
  \label{eq:fitting_line1}
\end{equation}
with $\alpha \simeq 0.8$ and $\beta \simeq -6$ 
that are obtained from our numerical results
by the least square fitting. 
The fitting line, eq.~(\ref{eq:fitting_line1}),
is expressed by thick solid lines in the plots.
For $\eta < 1$, the accretion rate declines, which  
corresponds to gap opening in the planetesimal disk.

\begin{figure}[tbp]
  \epsscale{1.00}	%
  \plottwo{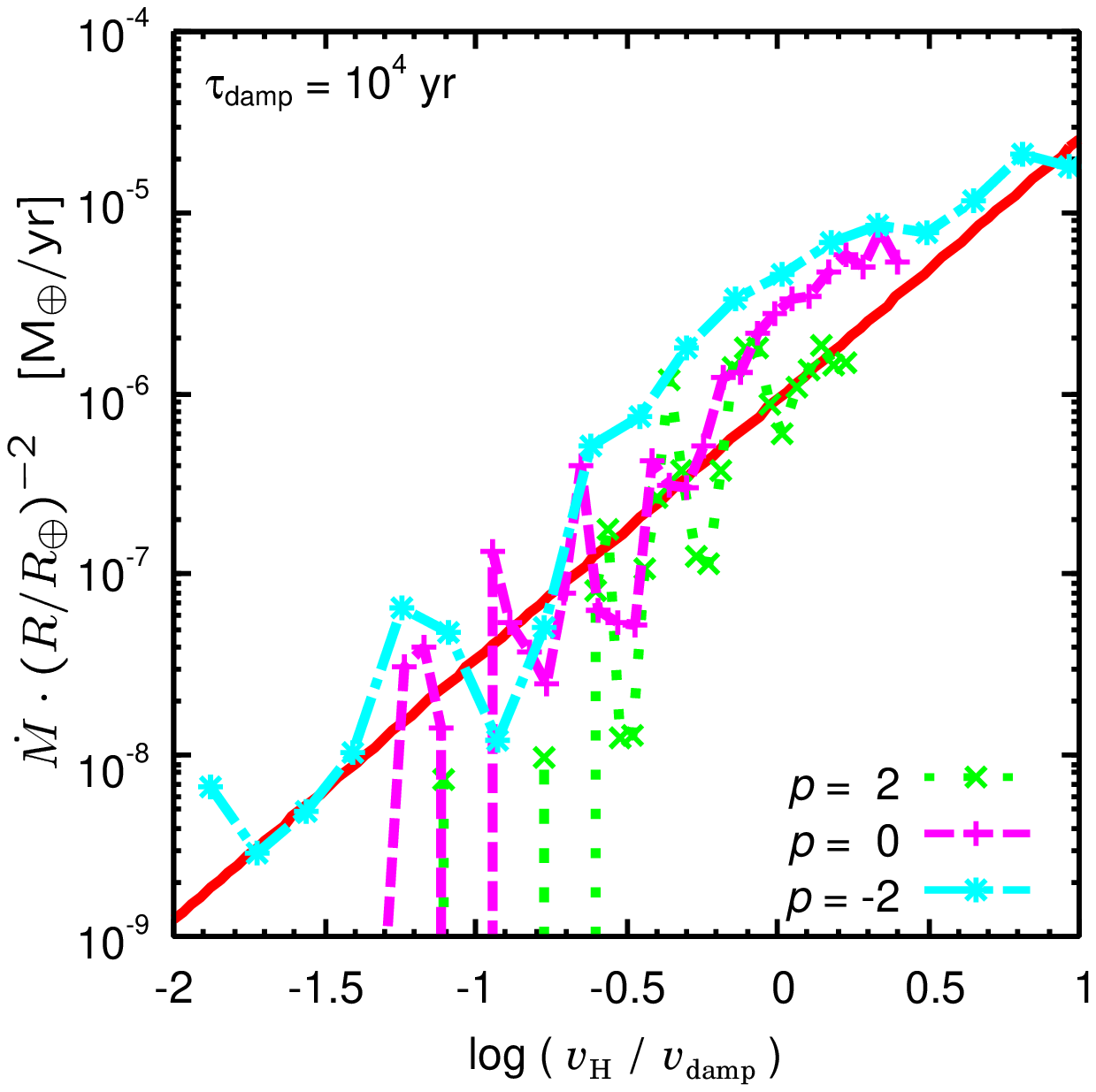}{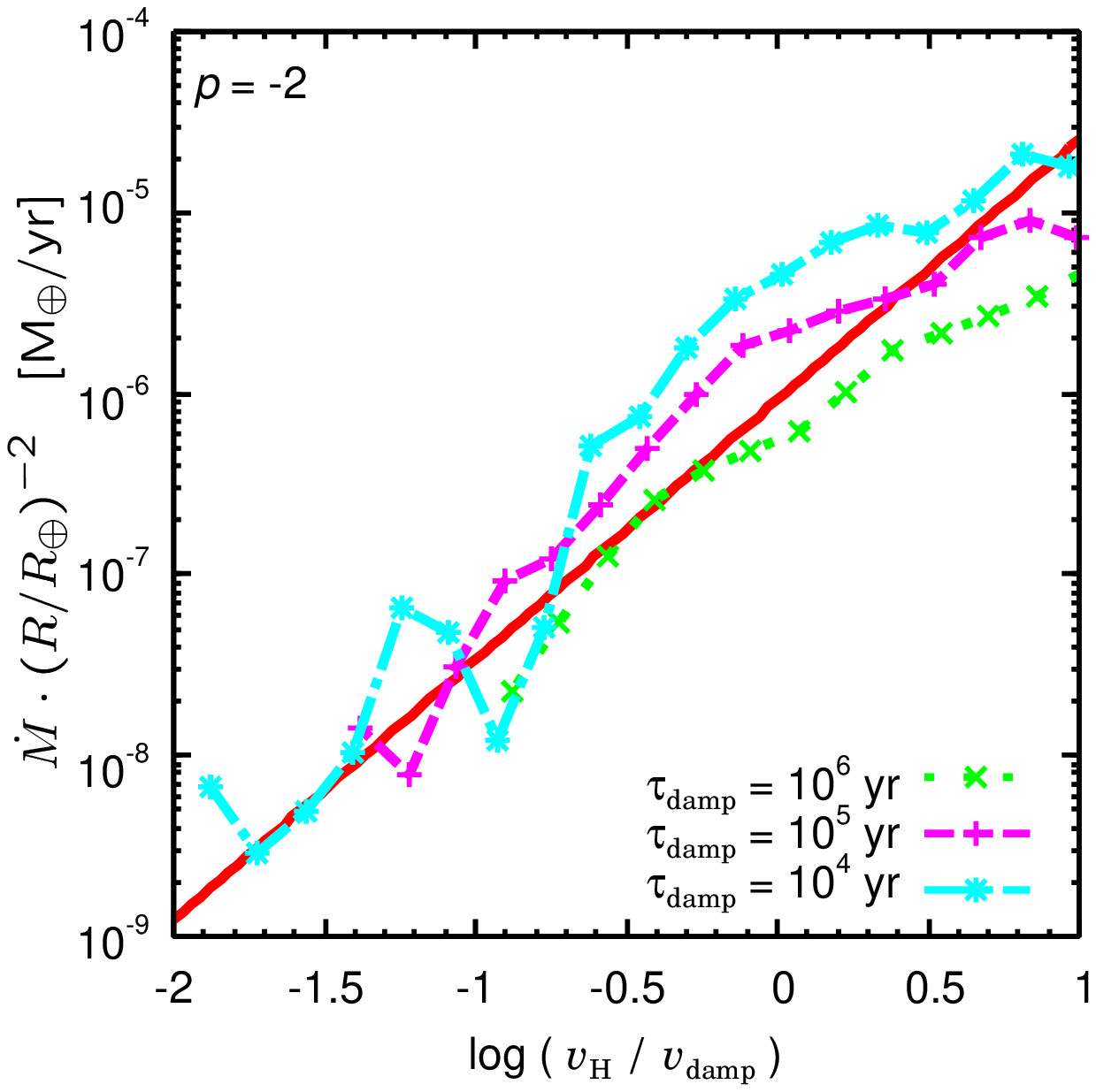}
  \caption{Evolution of the scaled planetesimal accretion rate 
as a function of $\eta  = v_H / v_{\rm damp}$ for $p = 2,0$ and $-2$ 
in the case of $\tau_{\rm damp} = 10^4$ yrs (left panel) and for 
$\tau_{\rm damp} = 10^6,10^5$ and $10^4$ yrs in the case of $p = -2$ (right panel).
The fitting formula, eq.~(\ref{eq:fitting_line2}),
is expressed by thick solid lines.}
  \label{fig:mlmp11_eta}
\end{figure}

Figure \ref{fig:mlmp11_eta} shows the evolution of 
the scaled planetesimal accretion rate as 
a function of $\eta = v_{\rm H}/v_{\rm damp}$. 
In the range of $\eta < 1$, the scaled planetesimal 
accretion rate is independent of planet accretion 
models with different $\dot{M}$--$M$ relations and
different $\tau_{\rm damp}$. 
This confirms that the accretion rate is
determined by $\eta$ for $\eta<1$.
From our numerical results, 
the planetesimal accretion rate in this case is given by
\begin{equation}
  \frac{dM_{\rm solid}}{dt} =
  10^{\beta} 
  \left( \frac{R}{R_\oplus}\right)^{2} 
  f_d \left( \frac{v_{\rm H}}{v_{\rm damp}} \right)^{\alpha} 
  \, \textrm{M}_\oplus \, \textrm{yr}^{-1} \, ,
  \label{eq:fitting_line2}
\end{equation}
with $\alpha \simeq 1.4$ and $\beta \simeq -6$. 

\begin{figure}[tbp]
  \epsscale{1.00}	%
  \plottwo{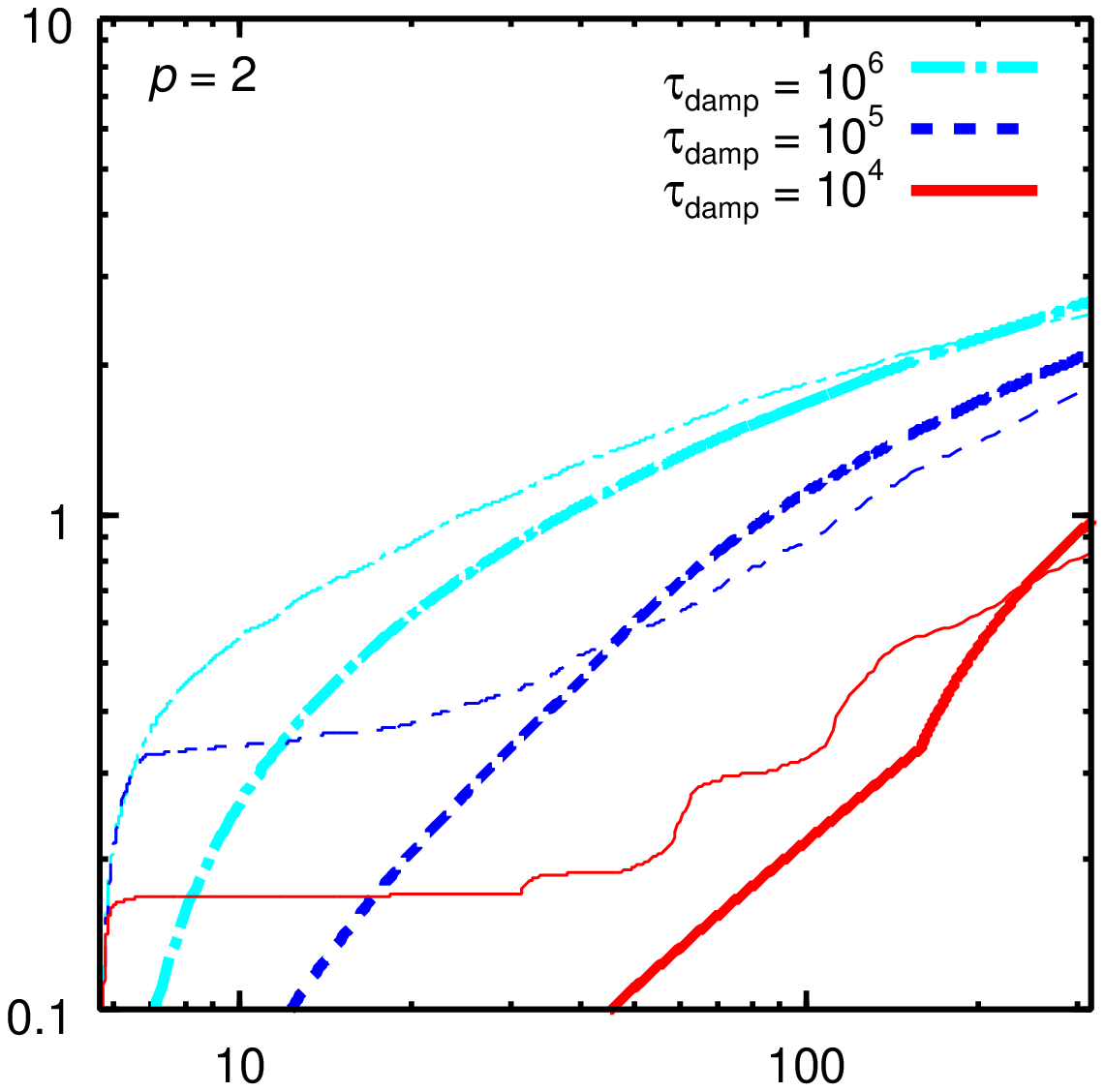}{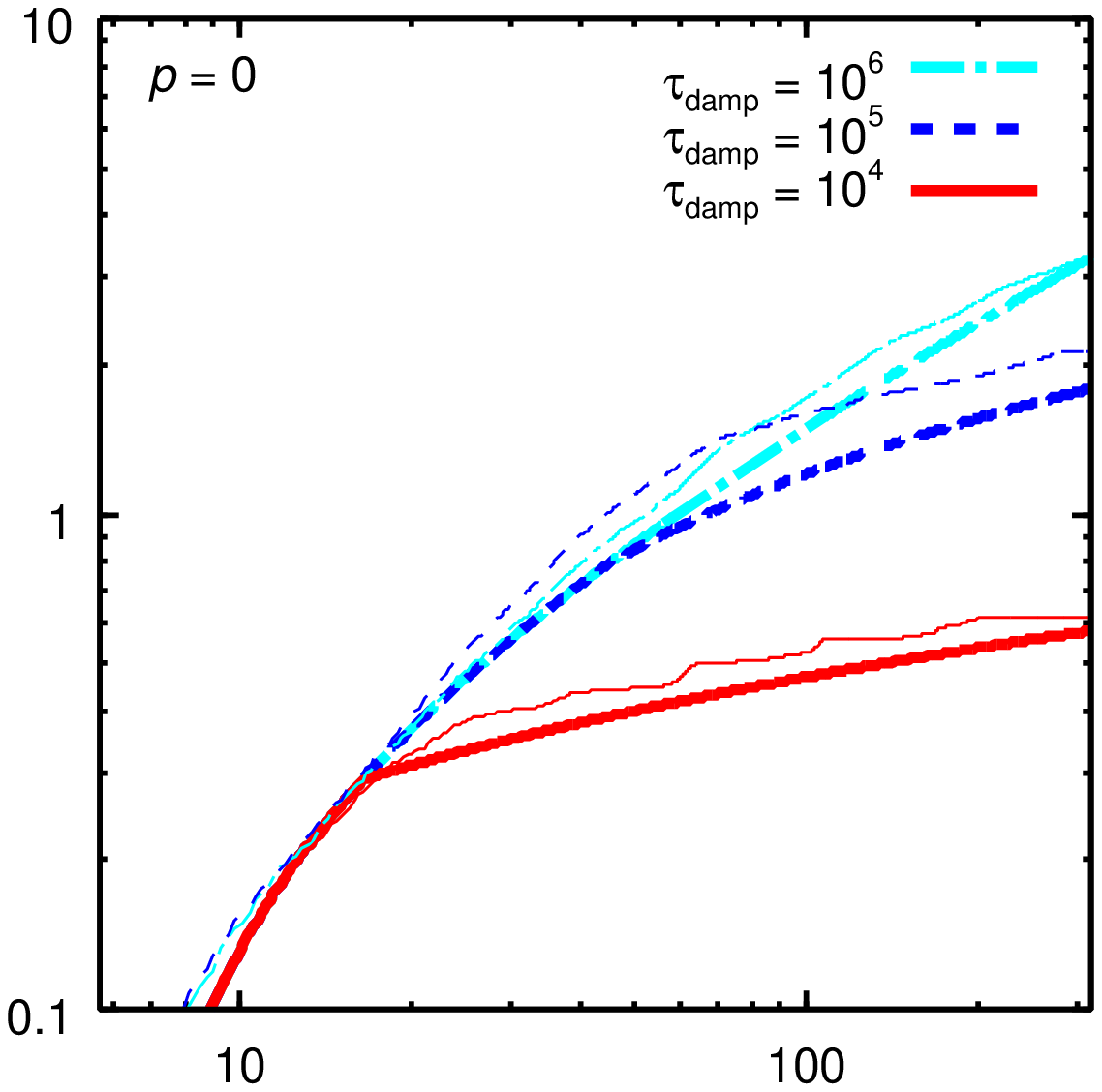}
  \caption{Comparison between the numerical simulations and the semi-analytical results.
  The left and right panels show the results
  for gas accretion with $p = 2$ and $p = 0$, respectively.
  The thin and thick curves represent 
  the numerical and semi-analytical results.}
  \label{fig:Arti_Compare_Amass}
\end{figure}

When the planet's mass has grown to $M$,
the total mass of planetesimals that infall in the envelope
($M_{\rm solid}(M)$) is
obtained by integrating
\begin{equation}
  \frac{dM_{\rm solid}}{dM} = \frac{dM_{\rm solid}}{dt} \frac{\tau_{\rm g,acc}}{M}
\label{eq:M_solid}
\end{equation} 
from 0 to $M$.
In figure~\ref{fig:Arti_Compare_Amass},
$M_{\rm solid}(M)$ evaluated by the above semi-analytical formulae
is compared with that obtained by orbital calculations
for individual gas accretion models in cases of
$\tau_{\rm damp} = 10^6$ yrs, $10^5$ yrs and $10^4$ yrs.
The semi-analytical formulae well reproduce the results of
orbital calculations except for early stages 
in which $M_{\rm solid}$ is so small that statistical fluctuation is large.
The formulae also reproduce numerical results
in their figure~9a in \citet{ZL07}.

\subsection{Application to Jupiter and Saturn}

In the preceding subsection, we investigated planetesimal accretion 
onto growing protoplanets with artificial gas accretion models 
and obtained semi-empirical formulae of the planetesimal accretion rate.
Applying this formulae to the more realistic gas accretion models 
in \S 2, we discuss the metallicity of envelopes of Jupiter and Saturn. 

\begin{figure}[tbp]
  \epsscale{1.0}	%
  \plottwo{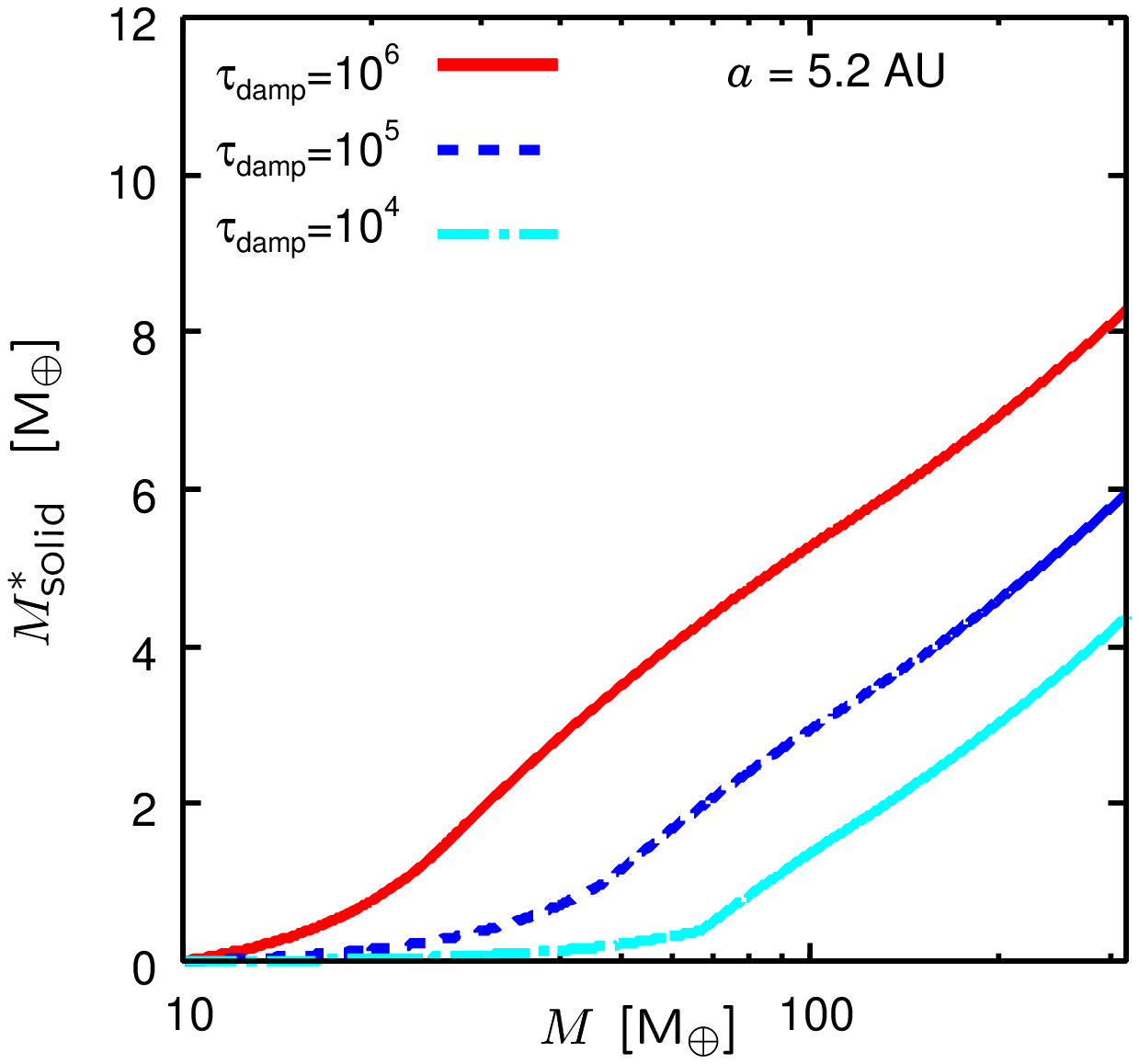}{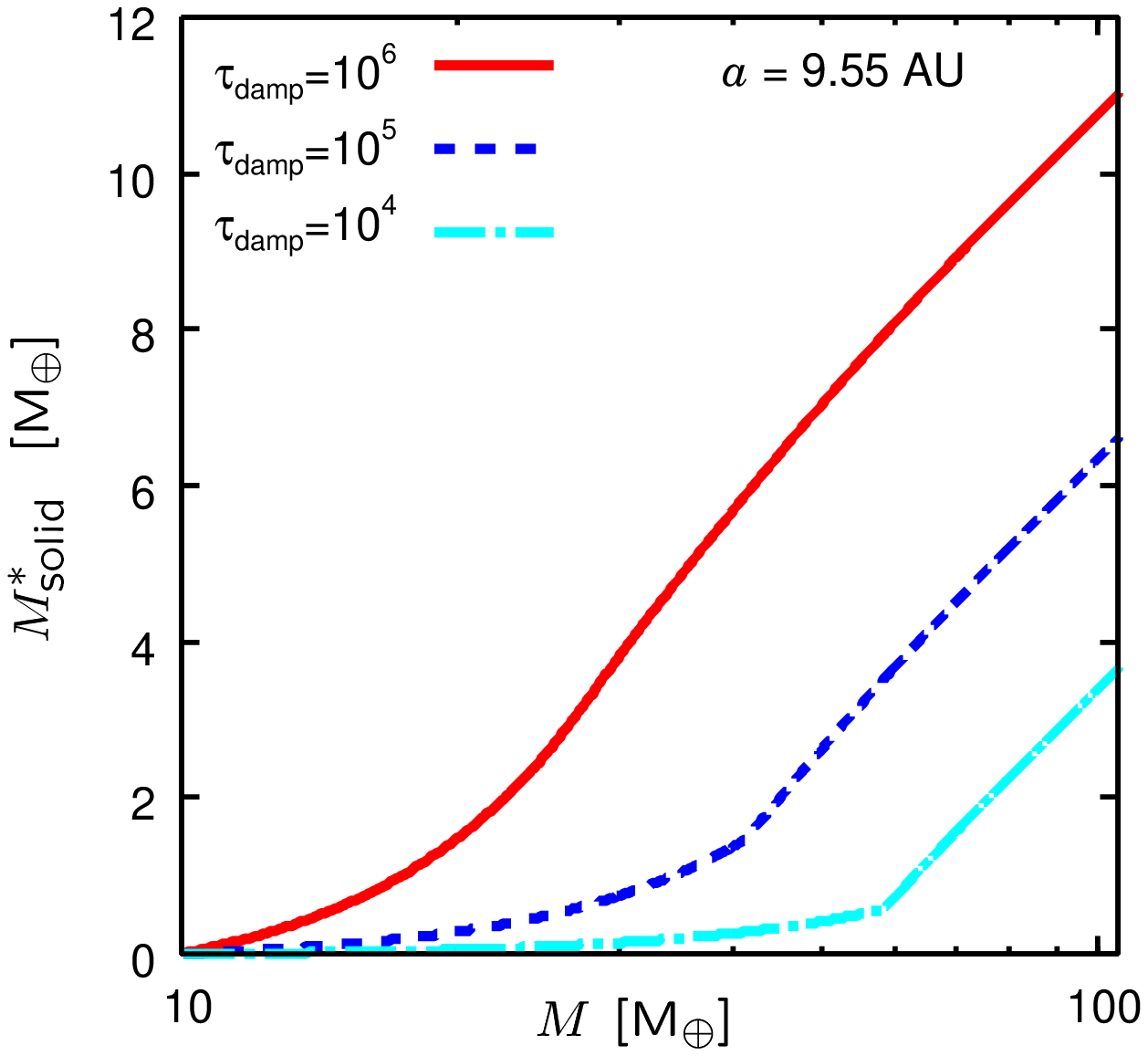}	%
  \caption{The evolution of cumulative mass of accreted planetesimals 
  as a function of $M$ in the case with $M_0 = 10M_\oplus$.
  The left and right panels show the results
  for $a = 5.2$ AU and $a = 9.55$ AU, which correspond to
  Jupiter and Saturn.
  Here, $R=2R_1$ and $f_d=2$ are assumed, where $R_1$ is
  the physical radius for mass $M$ and $\rho = 1$gcm$^{-3}$.
  For other $R$ and $f_d$, the accreted mass is
  multiplied by $(R/2R_1)^{1/2} (f_d/2)$.
}
  \label{fig:Amass_DAMP}
\end{figure}

Integrating eq.~(\ref{eq:M_solid}) with eq.~(\ref{eq:ikoma_model})
to $M_{\rm f}$, we estimate total mass of the accreted planetesimals 
in the cases of Jupiter ($M_{\rm f} = 318M_{\oplus}$,
$a_p = 5.2$ AU) and Saturn ($M_{\rm f} = 95M_{\oplus}$, $a = 9.55$ AU).
The evolution of $M_{\rm solid}$ is plotted in figure \ref{fig:Amass_DAMP}.
The three curves show 
the results with $\tau_{\rm damp}=10^4, 10^5$ and $10^6$ yrs.
It is likely that gas giants were inflated during 
gas accretion phase.
For a fixed $M$, $dM/dt \propto f_d \sqrt{\rho} R^2 \propto f_d \sqrt{R}$
(eq.~[\ref{eq:size-independent_accretion_rate}]).
In the figure, we plot the accreted planetesimal mass $M_{\rm solid}^*$
for $R=2R_1$ and $f_d=2$, where $R_1$ is
the physical radius for mass $M$ and $\rho = 1$gcm$^{-3}$.
For other $R$ and $f_d$, the accreted mass is
$M_{\rm solid} = (R/2R_1)^{1/2} (f_d/2) M_{\rm solid}^*$.

All the results show similar qualitative features of
evolution of $M_{\rm solid}$.
Planetesimal accretion is inhibited in early stages
by gap formation, but
rapid planetary growth due to gas accretion in later stages allows
planetesimal accretion.
With $\tau_{\rm damp}=10^6$ yrs, 
$M_{\rm solid} \simeq 6 (R/2R_1)^{1/2} (f_d/2) M_{\oplus}$
both for Jupiter and Saturn.
For shorter $\tau_{\rm damp}$, $M_{\rm solid}$ is smaller 
due to easier gap formation. 
We also did calculations starting from different core masses.
The resultant $M_{\rm solid}$ hardly changed, because
$dM_{\rm solid}/dM$ is negligibly small when
$M$ is small and gap is opened.
The amount of predicted $M_{\rm solid}$ can be
as large as that inferred from the internal structure model \citet{SG04},
if the planets are inflated and/or relatively large $f_d$ is
considered.

For the same $M$, $\rho$ and $f_d$, 
$M_{\rm solid}$ is larger for larger $a_p$.
Although Saturnian mass is $1/3$ of Jovian mass,
our model predicts that the mass of planetesimals 
falling into Saturnian envelope is comparable to
that into Jovian envelope.
More detailed internal structure models will
test our prediction.

\subsection{Phase 2}

So far, we have assumed that gas accretion immediately 
starts when $M_{\rm c}$ exceeds $M_{\rm c,hydro}$ 
without undergoing ``phase 2.''
In the previous subsection, we predicted
the planetesimal accretion rate as a function of planetary mass
based on the realistic gas accretion model.
With this accretion rate, we show that ``phase 2'' is not 
likely to occur. 

In the nominal model (J1 model) in \citet{PO96},
$f_d \simeq 2.5$, $a_p = 5.2$AU and $M_{\rm c} \simeq 10 M_{\oplus}$.
Then, they found that $\dot{M}_{\rm c} \simeq 10^{-6} M_{\oplus}/$yr is
maintained during ``phase 2'' with their maximally
efficient planetesimal accretion model.
As shown in eq.~(\ref{eq:crit_core_mass}),
this $\dot{M}_{\rm c}$ can marginally support gas envelope
around a $10 M_{\oplus}$ core.

First, we derive the condition for gap opening with
a realistic $\tau_{\rm g,acc}$ given by eq.~(\ref{eq:ikoma_model}).
Substituting eq.~(\ref{eq:ikoma_model}) into eq.~(\ref{eq:eta}),
\begin{equation}
\eta \simeq 0.8 \times 
10^{-6} \left(\frac{\tau_{\rm damp}}{10^{4} \textrm{yrs}}\right)^{1/2}
\left(\frac{M}{M_{\oplus}}\right)^{3.3} 
\left(\frac{a_p}{5\textrm{AU}}\right)^{3/4}.
\label{eq:eta2}
\end{equation}
With $a_p = 5.2$AU, $M \sim M_{\rm c} \sim 10 M_{\oplus}$ and
$\tau_{\rm damp} = 10^6$ yrs, we obtain $\eta \simeq 2 \times 10^{-2} \ll 1$.
Then, the gap should be opened up.
Our formula for $\eta < 1$ gives
\begin{equation}
\dot{M}_{\rm solid} \simeq 2.2 \times 10^{-6} f_d
\left( \frac{\rho_{\rm p}}{\rm 1gcm^{-3}} \right)^{-1/6}
\left( \frac{\tau_{\rm damp}}{10^4{\rm yrs}} \right)^{7/10}
\left( \frac{\tau_{\rm g,acc}}{10^4{\rm yrs}} \right)^{-7/5}
\left( \frac{M}{M_{\oplus}} \right)^{13/30} 
\left( \frac{a_p}{5{\rm AU}} \right)^{21/20} \ \rm M_{\oplus}/yr.
\label{mdot_gap}
\end{equation}
Substituting eq.~(\ref{eq:ikoma_model}) into this equation,
\begin{equation}
\dot{M}_{\rm solid} \simeq 0.9 \times 10^{-14} f_d
\left( \frac{\rho_{\rm p}}{\rm 1gcm^{-3}} \right)^{-1/6}
\left( \frac{\tau_{\rm damp}}{10^4{\rm yrs}} \right)^{7/10}
\left( \frac{M}{M_{\oplus}} \right)^{16/3} 
\left( \frac{a_p}{5{\rm AU}} \right)^{21/20} \ \rm M_{\oplus}/yr.
\label{mdot_gap2}
\end{equation}
For $\tau_{\rm damp} = 10^6$ yrs, $f_d \simeq 2.5$, $a_p = 5.2$AU 
and $M \simeq 10 M_{\oplus}$, 
$\dot{M}_c \simeq 1.1 \times 10^{-7} M_{\oplus}/$yr, 
which is one order smaller
than the planetesimal accretion rate that \citet{PO96} assumed.

\begin{figure}[tbp]
  \epsscale{0.6}	
  \plotone{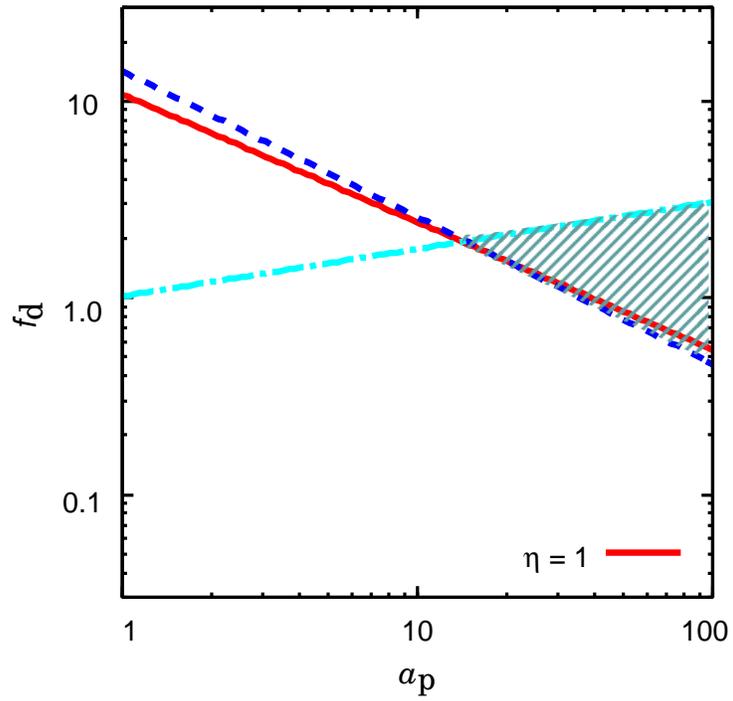}
  \caption{The parameter range in which
  phase 2 can occur, which is expressed by the shaded regions.
  The region above the solid line represents $\eta > 1$.
  Phase 2 can occur in the regions above the dashed line for $\eta < 1$ and
  in the regions below the dot-dashed line for $\eta > 1$.
}
  \label{fig:Phase2}
\end{figure}

We examine the possibility of phase 2 for other $f_d $ and $a_p$. 
For phase 2 to occur, $\dot{M}_{\rm c}$ must be maintained
to be as large as $\dot{M}$ for $M_{\rm c} \sim M_{\rm c,hydro}$.
Core mass can be approximately identified by core isolation mass
beyond the ice line \citep{KI98,KI02,IL04},
\begin{equation}
M_{\rm c,iso} \simeq 4.6 f_d^{3/2} \left(\frac{a_p}{5{\rm AU}}\right)^{3/4} M_{\oplus}.
\end{equation}
From eq.~(\ref{eq:crit_core_mass}) with the exponent derived by
assuming eq.~(\ref{eq:ikoma_model}), the accretion rate required by
occurrence of phase 2 is
\begin{equation}
\dot{M}_{\rm solid,2}
         \simeq 10^{-6} \left(\frac{M_{\rm c}}{10M_{\oplus}}\right)^{4.5} 
         M_{\oplus}/{\rm yr}
         \sim 3 \times 10^{-8} f_d^{6.75}
         \left(\frac{a_p}{5{\rm AU}}\right)^{3.4} M_{\oplus}/{\rm yr}.
\label{eq:mdot_phase_2}
\end{equation}
Substituting $M_{\rm c,iso}$ into $M$ in eq.~(\ref{eq:eta2}),
\begin{equation}
\eta \simeq 1.3 \times 10^{-4} f_d^5
\left(\frac{\tau_{\rm damp}}{10^{4} \textrm{yrs}}\right)^{1/2}
\left(\frac{a_p}{5\textrm{AU}}\right)^{13/4}.
\end{equation}
So, $\eta < 1$ is equivalent to
\begin{equation}
f_d < 3.8 \left(\frac{\tau_{\rm damp}}{10^{6} \textrm{yrs}}\right)^{-1/10}
\left(\frac{a_p}{5\textrm{AU}}\right)^{-13/20}.
\end{equation}
For this range of $f_d$ and $a_p > 3$AU (the ice line),
eq.~(\ref{mdot_gap}) with 
$M$ replaced by $M_{\rm c,iso}$ is always smaller than 
$\dot{M}_{\rm solid,2}$ given by eq.~(\ref{eq:mdot_phase_2}) 
(see figure \ref{fig:Phase2}).
For $\eta > 1$, on the other hand, 
\begin{equation}
\dot{M}_{\rm solid} \simeq 1.5 \times 10^{-5} f_d 
\left( \frac{\rho_{\rm p}}{\rm 1gcm^{-3}} \right)^{-1/6}
\left( \frac{\tau_{\rm g,acc}}{10^4{\rm yrs}} \right)^{-4/5}
\left( \frac{M}{M_{\oplus}} \right)^{2/5} 
\left( \frac{a_p}{5{\rm AU}} \right)^{6/5} \ \rm M_{\oplus}/yr.
\label{mdot_nogap}
\end{equation}
In the range of $f_d$ and $a_p$ that satisfy $\eta > 1$,
eq.~(\ref{mdot_nogap}) can reach $\dot{M}_{\rm solid,2}$ 
only at $a_p > 15$AU and $f_d \sim 1$, 
in which gas giant formation is unlikely
\citep{IL04}.
Thus, the predicted $\dot{M}_c$ never reaches the values 
required for phase 2.
We conclude that phase 2 is not likely to occur for formation of
giant planets.  
This conclusion is consistent with the ubiquity of
extrasolar gas giant planets.

\section{Conclusion}

We have investigated the planetesimal accretion rate onto growing giant planets 
through numerical simulations and analytical arguments.
The planet mass ($M$) is increased with assumed gas accretion rate onto the planet,
and orbits of planetesimals in the vicinity of the planet's orbit are
integrated with the effect of gas drag, but without self-gravity of the
planetesimals.

We first performed simulations with several different 
artificial gas accretion rates to clarify intrinsic physics determining
the planetesimal accretion rate.
A gap in the planetesimal disk is opened by a coupling effect of
gravitational scattering by the planet and gas drag damping.
Here, the gap formation means that most planetesimals are get out of the
feeding zone of the planet.
The scattering increases both $e$ and
$b$ keeping Jacobi energy constant, 
where $e$ is orbital eccentricity and $b$ is difference in semimajor axis
between the planet and the planetesimals.
Changes in $e$ and $bh$ are of the same order, where $h$ is 
reduced Hill radius defined by $(M/3M_\ast)^{1/3}$.
Since the gas drag predominantly damps $e$ after the scattering,
the gap is formed.
On the other hand, the width of the feeding zone is 
proportional to $h$.
Thus, the planet growth inhibits 
gap formation and competes with the scattering/damping process.

We derived the condition for the gap formation by
comparison between the eccentricity damping rate ($v_{\rm damp}$)
and the rate of expansion of the feeding zone due to the planet growth 
($v_{\rm H}$).
When $v_{\rm H}/v_{\rm damp} > 1$, the gap is not formed.
Then, the planetesimal accretion rate ($dM_{\rm solid}/dt$) is 
scaled by the ratio of the scattering rate 
$v_{\rm scat}$ to $v_{\rm H}$.
The numerical results are fitted as
\begin{equation}
  \frac{dM_{\rm solid}}{dt} =
  10^{-6} 
  \left( \frac{R}{R_\oplus}\right)^{2} 
  f_d \left( \frac{v_{\rm H}}{v_{\rm scat}} \right)^{0.8} 
  \, \textrm{M}_\oplus \, \textrm{yr}^{-1} \, ,
\end{equation}
where $R$ is physical radius of the planet and $f_d$ is
a scaling factor for surface density of the planetesimals
(eq.~[\ref{eq:surface_mass_density}]).
When the gap is formed ($v_{\rm H}/v_{\rm damp} < 1$), 
the accretion rate is significantly depleted.
We found that the accretion rate is scaled by 
$v_{\rm H}/v_{\rm damp}$ as
\begin{equation}
  \frac{dM_{\rm solid}}{dt} =
  10^{-6} 
  \left( \frac{R}{R_\oplus}\right)^{2} 
  f_d \left( \frac{v_{\rm H}}{v_{\rm damp}} \right)^{1.4} 
  \, \textrm{M}_\oplus \, \textrm{yr}^{-1} \, .
\end{equation}

Applying these formulae to the more realistic gas accretion models
described in \S 2, we found the followings:
\begin{enumerate}
\item In early stages when $M \sim O(10)M_{\oplus}$, a gap is opened
in the planetesimal disk. 
The planetesimal accretion rate is smaller than that required 
for phase 2 to occur.
This ensures efficient formation of gas giants, which may be consistent
with the ubiquity of extrasolar giant planets.
\item In later stages ($M \ga O(100)M_{\oplus}$),
the expansion of the feeding zone overwhelms the gap opening process,
so the gap is filled. 
Then, the planetesimal accretion becomes efficient.
\item The amount of infalling planetesimals into
the envelopes of Jupiter and Saturn in the late stages can be as large as
several $M_{\oplus}$, which may be consistent with
interior models for these planets.
\end{enumerate}
In this "realistic" model, we assumed that planetesimals
are infinitely supplied.  
However, if the accreted mass is significant, planetesimals distributed
in the regions inside isolated strong mean motion resonances can be consumed.
In that case, release of planetesimals from the resonance capture
by resonance overlapping due to planet mass increase may also become
a important factor \citep{ZL07}.

\citet{Guillot06} pointed out the correlation that
the amount of solid components of extrasolar transiting gas giants increases
with metallicity of their host stars that is proportional to $f_d$.
This trend is consistent with our formulae, because
$dM_{\rm solid}/dt \propto f_d$.
As this example shows, the analysis here 
will give deep insights into formation of 
extrasolar gas giants and their diversity.

\acknowledgments			%
This work is supported by JSPS.

\section*{Appendix}

The magnitude of $(e/h^2)$ in \S 4.3 is determined by 
a balance between damping due to the gas drag and 
excitation due to the planet's perturbations.
Since in the non-gap case, planetesimals are engulfed by the
feeding zone mainly through the parameter range of $(e/h) \la 1$,
we use eq.~(\ref{eq:scattering_velocity})
for definition of the parameter $\xi$.
However, gap opening is caused by
damping of relatively high orbital eccentricity, so 
we use the formula of excitation of
planetesimal eccentricity due to the protoplanet's 
perturbations for $(e/h) \ga 1$, in evaluating $(e/h)^2$.
Then the scattering timescale is given approximately by
Chandrasekahr's two-body scattering formula 
\citep[e.g.,][]{Stewart_Ida00,Ohtsuki02},
\begin{equation}
\tau_{e,{\rm scat}} \simeq 
\frac{1}{n_p \pi (GM/(e v_{\rm K})^2)^2 e v_{\rm K} \ln \Lambda},
\end{equation}
where $\ln \Lambda \sim 3$ and
$n_p$ is spatial density of the protoplanet, which is
given by inverse of volume of the planetesimal disk
in the feeding zone, 
$1/(2 \pi a_p \times 4\sqrt{3} h a_p)(e v_{\rm K}/\Omega_{\rm K})$.
Then, 
\begin{equation}
\tau_{e,{\rm scat}}
\simeq \frac{8\sqrt{3}\pi (e/h)^4}{27\pi} h^{-1} \frac{T_{\rm K}}{2\pi}
\simeq 1 \times 10^{2} (e/h)^4
       \left( \frac{M}{M_\oplus}\right)^{-1/3} 
       \left(\frac{a_p}{5\textrm{AU}} \right)^{3/2}  \textrm{yrs.} 
\end{equation}
From $\tau_{e,{\rm scat}} = \tau_{\rm damp}$, we obtain
\begin{equation}
(e/h)^2 \simeq 10 \left( \frac{\tau_{\rm damp}}{10^4 \textrm{yrs}} \right)^{1/2}
       \left( \frac{M}{M_\oplus}\right)^{1/6} 
       \left(\frac{a_p}{5\textrm{AU}} \right)^{-3/4}  \textrm{.} 
\label{eq:equili_e}
\end{equation}

\clearpage

\clearpage

\clearpage

\clearpage

\clearpage

\clearpage

\clearpage

\clearpage

\clearpage

\clearpage
\end{document}